\documentclass{article}

\usepackage[utf8]{inputenc}
\usepackage[english]{babel}
\usepackage[margin=1.3cm]{geometry}
\usepackage{mathtools}
\usepackage{multicol}
\usepackage{hyperref}

\usepackage{amsthm}
\usepackage{amssymb}
\usepackage{amsmath}

\title{On the Bogoliubov-Valatin transformation for fermionic Hamiltonians without a linear part}
\author{Davide Bonaretti \footnote{Università di Pisa, Pisa, Italy. ORCID:0009-0003-7771-9814}}
\newtheorem{theorem}{Theorem}
\newtheorem{lemma}{Lemma}
\newtheorem{requirement}{Requirement}

\newcommand\Tonde[1]{\, \left ( #1 \right )}
\newcommand\Quadre[1]{ \left [ #1 \right ]}
\newcommand\Graffe[1]{ \left \{ #1 \right \}}
\newcommand\Scalar[2]{ \ \left \langle #1,\, #2\, \right \rangle }
\newcommand{\Sum}[1]{ \sum_{#1}}

\newcommand\LowTrace[1]{ \mathrm{tr}\Graffe{ #1 }}

\newcommand\Transposed{	\mathrm{T} }
\newcommand\Enlarge{ \vphantom{\frac{1}{1}}}
\newcommand\Span[1]{ \mathrm{span}\Quadre{#1}}

\newcommand\Std{\ominus} 	%<--- This may be changed to modify the FunnySymbol 

\newcommand{\HL}[1]{\fbox{ \parbox{\linewidth}{#1}} }

\begin{document}

	\maketitle

	\begin{center}
		Abstract
	\end{center}
	\begin{quote}
		
		A self-contained treatment of the Bogoliubov-Valatin
		transformation for homogeneous fermionic Hamiltonians is presented. 
		The aim is to provide a quick reference that may also serve as supplementary
		material for a graduate-level course, and that can be understood 
		with quantum mechanics knowledge up to the level of the second quantization's rules.  
		The objective of the transformation is to cast a quadratic Hamiltonian into a diagonal form that
		resembles the Hamiltonian of a system of non-interacting particles.
		To obtain this, the first step consists in putting its coefficient matrix into its canonical form; 
		the transformation can always be performed on fermionic Hamiltonians, 
		only some care must be taken when this form is singular. 
		Having explained how to cast a general matrix into its standard form, a complete description of the 
		transformation is provided; a novel procedure is proposed here for the singular matrix case.  

	\end{quote} 
	\begin{center}
	\HL{
		\textbf{
		This article may be downloaded for personal use only. Any other use requires prior permission of the author and AIP Publishing. This article appeared in \textit{Am. J. Phys. 94, 239–244 (2026)} and may be found at 
		\url{https://doi.org/10.1119/5.0291967}}.\\

		Copyright (2025) Davide Bonaretti. This article is distributed under a Creative Commons Attribution-NonCommercial-NoDerivs 4.0 International (CC BY-NC-ND) License. \url{https://creativecommons.org/licenses/by-nc-nd/4.0/}
	}
	\end{center}

	\begin{multicols}{2} 

	\section{Introduction}
			The Bogoliubov-Valatin transformation is a procedure used to put quadratic Hamiltonians into a diagonal form 
	through the definition of new creation and destruction operators. 
	This approach seems to have first been used in a work of Holstein and Primakoff \cite{Holstein}, 
	but took its name by a paper of Bogoliubov
	\cite{Bog1947} on superfluidity, where it was applied to a Bose gas; later, a fermionic version 
	was developed by him to give a new interpretation of superconductivity 
	\cite{Bog1958} and independently exploited by Valatin \cite{Valatin1958}
	for the same purpose. In these works the transformation was applied to specific quadratic systems and 
	did not take the form of a general procedure.  
	A rather general treatment for the fermionic case 
	appeared in \cite{Lieb1961}\footnote{See Appendix A of \cite{Lieb1961}.};  
	a complete procedure for fermions can be found in the paper of Colpa \cite{Colpa1979}, 
	which also covers the case of quadratic Hamiltonians with a linear part. 

	This transformation unveils the fact that the eigendecomposition of a 
	quadratic Hamiltonian operator is a task far simpler than in the general case, because
	it can be rewritten to resemble the Hamiltonian of a system of non-interacting fermions or bosons;
	today it retains its historical place in the study 
	of superconductivity and superfluidity, but finds, for this reason, application 
	also as a general algebraic tool in the study of many-particle and  open quantum systems. 
	Indeed, fermionic and bosonic quadratic Hamiltonians 
	frequently appear in the approximation of mean field theories. 
	It is also well known that the dynamics of several spin models can be described 
	by fermionic quadratic Hamiltonians e.g. the 
	transverse-field Ising chain \cite{Pfeuty} and the XY model \cite{Lieb1961}.
	More recent cases of interesting fermionic quadratic systems 
	were provided by the Kitaev's models described in \cite{Kitaev2000}\footnote{
		In this work a Bogoliubov-Valatin transformation is also implicitly implied in
		the analysis of the toy model.
	} and \cite{Kitaev2006}.

	In this paper we are concerned only with the Bogoliubov-Valatin transformation for fermionic Hamiltonians
	without a linear part. 
	The purpose is to provide a complete and self-contained derivation of the procedure,
	which may be understood with the knowledge of quantum mechanics up 
	to the level of the second quantization's rules \cite{Landau}.
	The reader is supposed to be familiar with some topics in linear algebra,
	in particular with the theory of Hermitian products, 
	the theory of diagonalization, and the orthogonalization algorithms 
	\cite{Lang}.  

	Having introduced the notion of quadratic Hamiltonian and the details of the transformation,
	the conditions that have to be satisfied are pointed out.
	The transformation can always be obtained in the fermionic case and presents no particular difficulties, 
	however, some care is needed when the canonical coefficient matrix of the Hamiltonian is singular. 
	Having shown how to cast a general coefficient matrix into this standard form,
	a well-established method \cite{Colpa1979,VanHemmen}, 
	that is valid when the latter is invertible, is recovered and  
	a new and shorter procedure\footnote{See Appendix A and Appendix B of \cite{Colpa1979}.}
	for the singular matrix case is proposed. To better explain such an approach, 
	a simple numerical example is provided, 
	that shows step-by-step how to carry out the transformation when this situation occurs.

	\subsection{Notation}
	\begin{itemize}

	\item 
		Column vectors are written with bold letters. If not differently specified, they are assumed to be in 
		$\mathbb{C}^{2N}$ and all the matrices to be in $\mathbb{C}^{2N \times 2N}$. 
		Here $2N$ corresponds to the dimension of the Hamiltonian's coefficient matrix. 

	\item
		$I_N$ and $I_{2N}$ are the $N\times N$ and $2N\times 2N$ identity  matrices, respectively.

	\item
		To indicate the complex conjugation of numbers, vectors and matrices the symbol $*$ is used e.g. 
		$M^*$ is the complex conjugate of the matrix $M$. 

	\item
		The transposition of a matrix is denoted by the $\mathrm{T}$ symbol e.g. $M^\mathrm{T}$.

	\item
		The $\dagger$ symbol is adopted to denote the Hermitian conjugate of a matrix 
		i.e. given a matrix $M$, $M^\dagger := M^{*\mathrm{T}}$ is set,
		but it is also used to indicate the adjoint of operators that act on quantum states. 

	\item
		The scalar product between two vectors of $\mathbb{C}^{2N}$, say 
		$\mathbf{x}$ and  $\mathbf{y}$, is indicated as
		\begin{equation}
			\Scalar{ \mathbf{x}}{\mathbf{y}} := \sum_{k = 1}^{2N} x_k^*y_k
		\end{equation}
		and $\mathbf{x} \perp \mathbf{y}$ is written in place of $\Scalar{ \mathbf{x}}{\mathbf{y}} = 0$. 
		A similar writing is employed to indicate the orthogonality of two sets e.g.  $\mathcal{P} \perp \mathcal{Q}$
		means that for all $\mathbf{x} \in \mathcal{P}$ and $\mathbf{y} \in \mathcal{Q}$  is 
		$\mathbf{x} \perp \mathbf{y}$. 

	\item
		The Greek indexes take the values $1,...,2N$.
	\end{itemize}

	\section{Quadratic Hamiltonians}
				A homogeneous quadratic Hamiltonian operator has the form
			\begin{equation}
			\begin{array}{rl}
				\hspace{-1cm}\hat{H} := \displaystyle  \sum_{m,\,n = 1 }^N & \displaystyle \left ( \Enlarge \right.
						W_{mn} \hat{a}_m^\dagger \hat{a}_n
						+X_{mn} \hat{a}_m\hat{a}^\dagger_n  \\\\
				&  \hspace{1cm} \displaystyle +Y_{mn} \hat{a}_m \hat{a}_n +Z_{mn} \hat{a}_m^\dagger \hat{a}_n^\dagger \left ) \Enlarge \right.
			\end{array}
			\end{equation}
		where $\hat{a}_j$ are fermionic operators, that satisfy the anti-commutation
		relations
			\begin{equation}
				\Graffe{ \hat{a}_i, \hat{a}^\dagger_j} = \delta_{ij}
				\ \ \,
				\Graffe{ \hat{a}_i, \hat{a}_j} = 0
			.\end{equation}
		Adopting the notation of Nambu \cite{Nambu}\footnote{
			In this work Nambu rearranged fermionic field operators $\hat{\psi}(x)$ and 
			$\hat{\psi}^\dagger(x)$ into a tuple $\Psi(x) = (\hat{\psi}(x) , \hat{\psi}^\dagger(x) )$, 
			and this kind of notation remained associated to his name. 
		}, let us define the tuple
			\begin{equation}
				\hat{A} := ( \hat{a}_1, ..., \hat{a}_N, \hat{a}_1^\dagger, ..., \hat{a}_N^\dagger )
			,\end{equation}
		by means of which we can resume the fermionic properties in the formula
			\begin{equation}
				\Graffe{ \hat{A}_\mu, \hat{A}^\dagger_\nu } = \delta_{\mu\nu}
			.\end{equation}
		A matrix $H'$ can be conveniently chosen such that the Hamiltonian is cast in the form
			\begin{equation}
				\hat{H} = \sum_{\mu,\,\nu = 1}^{2N} H'_{\nu\mu} \hat{A}_\mu \hat{A}_\nu^\dagger 
				\label{eq:HQuadGen}
			.\end{equation}
		The matrices $W,X,Y,Z$, or equivalently $H'$, must ensure that $\hat{H} = \hat{H}^\dagger$. 
		Assuming this, the Hermitian part\footnote{A valid coefficient matrix
		can be non-Hermitian as well. A proof is provided at the end of Section 5.}
		of the matrix $H'$ would be the only one contributing in the
		Eq.\,(\ref{eq:HQuadGen}), as we could rewrite
			\begin{equation}
				\hat{H} = \frac{\hat{H}+\hat{H}^\dagger}{2} 
					= \sum_{\mu,\,\nu } H_{\nu\mu} \hat{A}_\mu \hat{A}_\nu^\dagger 
				\label{eq:HQuad}
			\end{equation}
		with $H := (H'+{H'}^\dagger)/2$.
		Therefore we can characterize a quadratic Hamiltonian 
		with a Hermitian coefficient matrix $H$. 
		Conversely, it can be verified that assuming a Hermitian 
		coefficient matrix grants the Hermiticity of its associated Hamiltonian operator.

	\section{Purpose of the transformation}
				We want to build new fermionic operators as a linear combination
		of the $\hat{A}_\mu$ operators, in the reference of which $\hat{H}$
		takes a diagonal form and acts like the Hamiltonian of a set of non-interacting free fermions.
		More precisely, let
			\begin{equation}
				\hat{B}_\mu := \sum_{\mu'} U_{\mu \mu' } \hat{A}_{\mu'}
				\label{eq:BAConnection}
			;\end{equation}
		we require that a form
			\begin{equation}
				\hat{H} = c + \sum_{\mu'} d_{\mu'} \hat{B}_{\mu'} \hat{B}_{\mu'}^\dagger 
				\label{eq:FinalForm}
			\end{equation}
		is obtained,  where 
		$\hat{B} = \Tonde{ \hat{b}_1, ..., \hat{b}_N, \hat{b}^\dagger_1, ..., \hat{b}_N^\dagger}$ is a tuple
		of new fermionic operators satisfying
			\begin{equation}
				\Graffe{ \hat{B}_\mu, \hat{B}_\nu^\dagger } = \delta_{\mu\nu}
			.\end{equation}
		Putting these operators in their normal order the Hamiltonian could be rewritten as
			\begin{equation}
				\hat{H} =  \sum_{j = 1}^{N} ( d_{j+N} - d_{j} ) \hat{b}_j^\dagger \hat{b}_j + 
					\sum_{j = 1}^{N} d_j + c
				\label{eq:OrderedFinalForm}
			.\end{equation}
		To enable this form possibly to represent the Hamiltonian of a set of non-interacting
		fermions, we also have to require that the vacuum state of the fermionic operators 
		$\hat{b}_j$ corresponds to a ground state of $\hat{H}$.
		We will see, when discussing the main theorem, that the number $c$ is obtained when 
		casting the coefficient matrix into its standard form, while the $d_\mu$ correspond to
		the eigenvalues of the latter.

	\section{Properties of the matrix $U$}
			The requests made in the previous section impose some restrictions on the matrix $U$. 

	Let 
	\begin{equation}	
		\Omega := \Tonde{ \begin{array}{cc}
					0	&	I_N	\\	
					I_N 	&	0 	\\
				\end{array}} 
		;\end{equation}
	it can be seen that $\Omega^2 = I_{2N}$. 
	By construction
		\begin{equation}
			 \hat{A}^\dagger_\mu= \Sum{\mu'} \Omega_{\mu\mu'} \hat{A}_{\mu'}
		\end{equation}
	and
		\begin{equation}
			 \hat{B}^\dagger_\mu= \Sum{\mu'} \Omega_{\mu\mu'} \hat{B}_{\mu'}
		.\end{equation}
	The matrix $U$ must satisfy
			\begin{equation}
				\hat{B}_\mu^\dagger = \Tonde{ \Sum{\mu'} U_{\mu\mu'} \hat{A}_{\mu'}}^\dagger 
						= \Sum{\mu'\nu'} U^*_{\mu \mu'} \Omega_{\mu'\nu'} \hat{A}_{\nu'}
				\label{eq:QConditionOrigin}
			.\end{equation}
	Renaming $Q :=  U - \Omega U^* \Omega$,  Eq. (\ref{eq:QConditionOrigin}) can be rewritten as
			\begin{equation}
				\Sum{\mu'} Q_{\mu \mu'} \hat{A}_{\mu'} = 0
			.\end{equation}
	Thus, for all $\hat{A}^\dagger_\nu$, it holds
			\begin{equation}
				\Sum{\mu'} Q_{\mu \mu'} \Tonde{ \hat{A}_{\mu'} \hat{A_\nu}^\dagger 
						+ \hat{A}_\nu^\dagger \hat{A}_{\mu'} } = 0
			.\end{equation}
	However, $\hat{A}_{\mu'} \hat{A_\nu}^\dagger + \hat{A}_\nu^\dagger \hat{A}_{\mu'} \equiv 
		\Graffe{ \hat{A}_{\mu'}, \hat{A}^\dagger_{\nu}} = \delta_{\mu'\nu}$, implying
			\begin{equation}
				Q_{\mu \nu } = 0
			,\end{equation}
	and for this reason it must be
			\begin{equation}
				U = \Omega U^* \Omega
				\label{eq:UFirstReq}
			.\end{equation}
	Defining the anti-linear function 
		\begin{equation}
			J( \mathbf{x}) :=  \Omega \mathbf{x}^*
		,\end{equation}
	the condition in Eq. (\ref{eq:UFirstReq}) implies that the matrix $U$ is of the form \cite{VanHemmen}:
	\begin{requirement} $ \boxed{U = \Tonde{ \Enlarge
					\mathbf{x_1}, ..., \mathbf{x_N}, 
					J(\mathbf{x_1}), ..., J(\mathbf{x_N})} }$ \\
	\end{requirement}
	\noindent for some $\mathbf{x}_1,...,\mathbf{x_N}$ in $\mathbb{C}^{2N}$. \\

	Recalling that
			\begin{equation}
				\Graffe{ \hat{B}_\mu, \hat{B}_\nu^\dagger } 
						= \Sum{\mu'\nu'} U_{\mu\mu'} U^\dagger_{\nu'\nu} 
							\Graffe{ \hat{A}_{\mu'}, \hat{A}^\dagger_{\nu'}}
						= \delta_{\mu\nu}
			,\end{equation}
	we obtain the
	\begin{requirement}
		$\boxed{ UU^\dagger = I_{2N} }.$
	\end{requirement}
		
	Let us recall now the Eq. (\ref{eq:OrderedFinalForm}). To ensure that the vacuum state of the $\hat{b_j}$ 
	operators is a ground-state of $\hat{H}$ it is necessary and sufficient to impose the
	\begin{requirement}
		 $\boxed{d_{j+N} \geq d_j \mathrm{\ for\ } j = 1,...,N }$
	\end{requirement}
	\noindent Note that if $d_{j+N} = d_{j}$ for some $j$, then there would be many ground states. The vacuum state
	would be one of them.

	\section{Constructing the transformation}
			
	In this section some useful lemmas are provided, then the main theorem is proven, explaining
	how to obtain the transformation.
	
	The first step of the procedure is casting the coefficient matrix into its standard form, 
	which in the following is denoted with the symbol $\Std$; its definition is provided in Lemma 1.
	Both Lemmas 1 and 2 point out some of its properties, providing the main arguments that 
	will lead to the final diagonal operator of Eq. (\ref{eq:FinalForm}).  
	In Lemma 3 a function is described, that will be employed to handle the case when 
	the standard form is singular.

		\begin{lemma}
		Let $M$ be a Hermitian matrix,
			\begin{equation}
				 M_\Std := \frac{1}{2} \Tonde{ M  - \Omega M^\Transposed \Omega }
				\label{eq:StandardForm}
			,\end{equation}
		and
			\begin{equation}
				\Graffe{ J, M }(\mathbf{x}) := J( M \,\mathbf{x}) + M J (\mathbf{x})
			.\end{equation}
		It holds 
			\begin{equation}
				\Graffe{ J, M} = 0\ \ \Longleftrightarrow\ \ M_\Std = M.
			\end{equation}
		\label{lemma:First}
	\end{lemma}
	\begin{proof}

		$M$ being Hermitian means $M^T = M^*$, hence for all $\mathbf{x}$ it holds
			\begin{equation}
				J( M J( \mathbf{x})) = (\Omega M \Omega \,\mathbf{x}^*)^* 
					= \Omega M^\Transposed \Omega \,\mathbf{x}
			.\end{equation}
		Substituting this equivalence into Eq.\,(\ref{eq:StandardForm})  we get
			\begin{equation}
				 M_\Std \,\mathbf{x} = \frac{1}{2} \Tonde{ \Enlarge M\mathbf{x}  - J(M J(\mathbf{x}))}
			.\end{equation}
		Exploiting the identity $J \circ J = \mathrm{Id}$, we can see that
			\begin{equation}
				\Graffe{ J, M }(\mathbf{x}) = 2( M - M_\Std ) J( \mathbf{x} )
			,\end{equation}
		thus we have have proven the lemma.
	\end{proof}\vspace{0.4cm}
	\begin{lemma}
		Let $M$ be a Hermitian matrix and $\mathcal{B}(M)$ a complete orthonormal basis for $\mathbb{C}^{2N}$ 
		made of eigenvectors of $M$. 
		If $\mathbf{x}$ is an eigenvector of $M$ we denote its eigenvalue as $\lambda_M(\mathbf{x})$.
		Let
			\begin{equation}
				\mathcal{N}_M := \Graffe{\, \mathbf{x} \in \mathcal{B}(M)  \ :\ \lambda_M(\mathbf{x}) < 0 }
			\end{equation}
		and
			\begin{equation}
				\mathcal{K}_M := \Graffe{\, \mathbf{x} \in \mathcal{B}(M)  \ :\ \lambda_M(\mathbf{x}) = 0 }
			.\end{equation} \\
		
		\noindent If $\Graffe{ J, M } = 0$, then
		\begin{enumerate}
		\item
			The kernel\,\footnote{It corresponds to $\Span{\mathcal{K}_M}$.} of $M$  is 
			invariant under the action of J.
		\item
			$\mathcal{N}_M\,\cup\,\mathcal{K}_M\,\cup\,J(\mathcal{N}_M)$ 
			is a complete orthonormal basis for $\mathbb{C}^{2N}$.
		\end{enumerate}
		\label{lemma:Second}
	\end{lemma}
	\begin{proof}
		Let $\Graffe{ J, M } = 0$. 
		If $\mathbf{x}$ in an eigenvector of $M$ then
			\begin{equation}
			 	MJ(\mathbf{x})+ \lambda_M(\mathbf{x}) J( \mathbf{x} ) = 0
			,\end{equation}
		implying that $J(\mathbf{x})$ is an eigenvector of $M$ with eigenvalue $-\lambda_M(\mathbf{x})$.

		\begin{enumerate}
		\item
			If $\mathbf{x}$ is in the kernel of $M$ then $\lambda_M(J(\mathbf{x}))=0$ and  $J(\mathbf{x})$
			belongs to it, too.
		\item
			It can be seen that for all vectors $\mathbf{y},\,\mathbf{z}$ it holds
				\begin{equation}
					\Scalar{ \mathbf{y} }{ \mathbf{z}} = \Scalar{ J(\mathbf{y}) }{ J(\mathbf{z})}^*
				.\end{equation}
			Thus, as the vectors in $\mathcal{N}_M$ are orthonormal, the vectors in
			$J(\mathcal{N}_M)$ are orthonormal too. 

			Note that $\mathbf{x} \perp J(\mathbf{x})$ for any $\mathbf{x}\in \mathcal{N}_M$, 
			because $M$ is Hermitian, and $\mathbf{x}$ and $J(\mathbf{x})$ have different eigenvalues. 
			This implies $\mathcal{N}_M \perp J(\mathcal{N}_M)$.
			Similarly, we have $\mathcal{N}_M \perp \mathcal{K}_M$ and $J(\mathcal{N}_M) \perp \mathcal{K}_M$, 
			thus we can state that
			$\mathcal{B}'(M) := \mathcal{N}_M\,\cup\,J(\mathcal{N}_M) \cup \mathcal{K}_M$ is a set of orthonormal
			eigenvectors of $M$. 

			If this were not a complete basis for $\mathbb{C}^{2N}$, then there would be an $\mathbf{x} \in
			\mathcal{B}(M)$ such that $\mathbf{x} \notin \Span{ \mathcal{B}'(M)}$. 
			The only possibility would be that $\mathbf{x}$ is an eigenvector of $M$ 
			with $\lambda_M(\mathbf{x})>0$. But in this case $J(\mathbf{x}) \in \Span{\mathcal{N}_M}$ and
			consequently $\mathbf{x} \in \Span{J(\mathcal{N}_M)}$, which would be a contradiction because
			$\Span{J(\mathcal{N}_M)} \subseteq \Span{ \mathcal{B}'(M)}$. 
			
		\end{enumerate}
	\end{proof}
	\begin{lemma}
		Let 

			\begin{equation}
				L(\mathbf{v}) := \begin{cases}
						i\mathbf{v}  & \mathrm{if\ } J(\mathbf{v}) = -\mathbf{v} \\
						\mathbf{v}+J(\mathbf{v})   & \mathrm{otherwise}  \\
				\end{cases} 
			.\end{equation}

		We can state that
		\begin{enumerate}
		\item
			$J(\,L(\mathbf{v})) = L(\mathbf{v})$.
		\item
			If $\mathbf{v} \neq 0$ then $L(\mathbf{v}) \neq 0$.
		\item
			If $J(\mathbf{u}) = \mathbf{u}$ and $\mathbf{u}\perp \mathbf{v}$ then 
			$\mathbf{u} \perp  L(\mathbf{v}) $.
		\end{enumerate}

		\label{lemma:Third}
	\end{lemma}
	\begin{proof} \noindent
		\begin{enumerate}
		\item 
			This is evident from the definition.
		\item 
			Assume  $\mathbf{v} \neq 0$ with $L(\mathbf{v}) = 0$, then
			$L(\mathbf{v}) \neq i\mathbf{v}$, which means
			$L(\mathbf{v}) = \mathbf{\mathbf{v}} + J( \mathbf{v}) =0 $, implying $L(\mathbf{v}) = i\mathbf{v}$, 
			which is a contradiction.
		\item
			If $L(\mathbf{v}) = i\mathbf{v}$ the third statement clearly holds. Otherwise, 
			if $L(\mathbf{v}) = \mathbf{v} + J(\mathbf{v})$, we have
			$\Scalar{ \mathbf{u}}{ L(\mathbf{v})} = \Scalar{ \mathbf{u}}{J(\mathbf{v})} 
			= \Scalar{\mathbf{v}}{J(\mathbf{u})} = \Scalar{\mathbf{v}}{\mathbf{u}} = 0 $.
		\end{enumerate}
	\end{proof}

	\begin{theorem}
		For each $\hat{H}$ we can find a matrix $U \in \mathbb{C}^{2N\times 2N}$ such that $\hat{H}$ takes
		the diagonal form of Eq. (\ref{eq:FinalForm}) and the Requirements 1, 2, and 3
		are satisfied. In other words, we can always recombine the original fermionic operators
		in such a way that $\hat{H}$ resembles the Hamiltonian of a system of non-interacting
		fermions with $N$ modes.
	\end{theorem}
	\begin{proof}
%		Below we construct this transformation matrix $U$.

		Let 
			\begin{equation}
				\hat{G}_{\mu\nu} := \hat{A}_\mu\hat{A}^\dagger_\nu
			.\end{equation}
		The Hamiltonian can then be rewritten in the compact form
		\begin{equation}
			\hat{H}  = \LowTrace{H\hat{G}}
			\label{eq:CompactH}
		.\end{equation}
		The matrix of operators $\hat{G}_{\mu\nu}$ has the property

			\begin{equation}
				\hat{G}  =  I - \Omega \hat{G}^\Transposed \Omega
			.\end{equation}
		Inserting this equivalence into Eq.\,(\ref{eq:CompactH}) and exploiting the 
		properties of the trace operation we obtain

			\begin{equation}
				\hat{H} =  - \LowTrace{ \Omega H^\Transposed \Omega \hat{G} } + \LowTrace{H}
			,\end{equation}
		thus, it holds
			\begin{equation}
			\begin{array}{l}
				\hat{H} = \frac{1}{2} \Tonde{ \Enlarge \LowTrace{ H \hat{G} } 
					- \LowTrace{ \Omega H^\Transposed  \Omega \hat{G} } + \LowTrace{H}} \\\\
				\hspace{0.5cm} = \LowTrace{H_\Std \hat{G}} + \frac{1}{2} \LowTrace{H}
				\label{eq:Decomposition}
			\end{array}
			.\end{equation}

		Due to Lemmas \ref{lemma:First} and \ref{lemma:Second},  
		$\mathcal{N}_{H_\Std} \cup \mathcal{K}_{H_\Std} \cup J( \mathcal{N}_{H_\Std})$ 
		is  an orthonormal basis of eigenvectors of $H_\Std$. \\

		If $H_\Std$ is invertible, let
			\begin{equation}
				U = \Tonde{ \vphantom{\frac{1}{1}}
					\mathbf{x}_1, ..., \mathbf{x}_N,  
					J(\mathbf{x}_1), ..., J(\mathbf{x}_N)  
				}
			\label{eq:InvertibleCaseU}
			\end{equation}
		where $\mathbf{x}_1, ...,\mathbf{x}_N \in \mathcal{N}_{H_\Std}$.

		Otherwise, if $H_\Std$ is singular, let $\mathbb{K}$ be its kernel. 
		To handle this special case, we first find a basis of
		$\mathbb{K}$ made of J-invariant orthogonal vectors. 
		This is possible because $\mathbb{K}$ is invariant under the action of $J$. 
		We pick an initial non-null vector $\mathbf{v_1}$ of $\mathbb{K}$ and let
			\begin{equation}
				\mathbf{y}_1 := 	L(\mathbf{v_1})
			.\end{equation}
		We proceed recursively: assume that we have built a set of orthogonal vectors up to an index $m$,
		namely $\Graffe{ \mathbf{y}_1, ..., \mathbf{y}_m}$.  	
		Take a non-null vector $\mathbf{v}_{m+1}$ of $\mathbb{K}$ 
		such that, 
			\begin{equation}
			\begin{array}{lr}
				\mathbf{v}_{m+1} \perp \mathbf{y}_j	& \mathrm{for\ }j= 1, ..., m  \\
			\end{array}
			.\end{equation}
		Lemma \ref{lemma:Third} guarantees that $L(\mathbf{v}_{m+1}) \neq 0 $ and
			\begin{equation}
			\begin{array}{lr}
				L(\mathbf{v}_{m+1}) \perp \mathbf{y}_j	& \mathrm{for\ }j= 1, ..., m  \\
			\end{array}
			\end{equation}
		and then we define the new orthogonal vector to be
			\begin{equation}	
				\mathbf{y}_{m+1} := L( \mathbf{v}_{m+1})
			.\end{equation}
		The procedure is repeated until a complete basis for $\mathbb{K}$ is obtained.

		By construction $\mathbb{K}$ is of even dimension, namely $2l$;
		this is due to Lemma 2, which ensures that $\mathcal{K}_{H_\ominus}$ contains
		an even number of linearly independent vectors.  Thus, we can define the linear function

		\begin{equation}
			\tilde{R} := \sum_{j=1}^l i \Tonde{ \vphantom{\frac{1}{1}}
					\mathbf{y}_{j} \Scalar{ \mathbf{y}_{j+l}}{ \cdot}
					-
					\mathbf{y}_{j+l} \Scalar{\mathbf{y}_{j}}{ \cdot}}
		.\end{equation}
		We can observe that 

			\begin{equation}
				J \circ \tilde{R}  + \tilde{R} \circ J = 0 
				\label{eq:RJAnticommute}
			.\end{equation}
		Let $R$ be the matrix associated to $\tilde{R}$ in the reference of the canonical basis of $\mathbb{C}^{2N}$. 
		$R$ is Hermitian and because of Eq.\,(\ref{eq:RJAnticommute}) 
			\begin{equation}
				\Graffe{ J, R } = 0
			.\end{equation}
		By construction,  as a function from $\mathbb{K}$ to $\mathbb{K}$,  $\tilde{R}$ is invertible, 
		while it is null for all vectors that are orthogonal to $\mathbb{K}$.
		Thus, invoking  again Lemma \ref{lemma:Second} we get that
		the set $\mathcal{N}_{R} \cup J( \mathcal{N}_{R} )$ is a complete orthonormal basis of $\mathbb{K}$.
		Consequently
			\begin{equation}
				\mathcal{N}_{H_\Std} \cup \mathcal{N}_{ R} \cup 
				J(  \mathcal{N}_{ H_\Std} \cup \mathcal{N}_{ R} )
			\end{equation}
		is a complete orthonormal basis for $\mathbb{C}^{2N}$ made of eigenvectors of $H_\Std$.
		Hence for singular $H_\Std$ we define the transformation matrix as
			\begin{equation}
				U = \Tonde{ \Enlarge 
					\resizebox{0.7\linewidth}{0.25cm}{
						$\mathbf{x}_1, ..., \mathbf{x}_n,  
						\mathbf{k}'_1, ..., \mathbf{k}'_l, 
						J(\mathbf{x}_1), ..., J(\mathbf{x}_n),
						J(\mathbf{k}'_1), ..., J(\mathbf{k}'_l)  $
					} }
			\label{eq:SingularCaseU}
			\end{equation}
		where $\mathbf{x}_1,...,\mathbf{x}_n \in \mathcal{N}_{H_\Std}$ and 
		$\mathbf{k}'_1,...,\mathbf{k}'_l \in \mathcal{N}_R$.\\\\

		In both cases, Eq. (\ref{eq:InvertibleCaseU}) and Eq. (\ref{eq:SingularCaseU}),  
		the matrix $U$ satisfies Requirements 1, 2, and diagonalizes 
		the matrix $H_\Std$. Let the $d_\mu$ correspond to its eigenvalues, taken according
		to the order of the column vectors in $U$, read from left to right. 
		Substituting the inverse of the relationship in Eq. (\ref{eq:BAConnection})
		into Eq. (\ref{eq:Decomposition}) we get 

			\begin{equation}
			\begin{array}{rl}
				\hat{H} =& \displaystyle \sum_{\mu'\nu'} 
					( U^\dagger H_\Std U)_{\mu'\nu'} \hat{B}_{\nu'} \hat{B}^\dagger_{\mu'}
					+ \frac{1}{2}\LowTrace{H} \\\\
					=& \displaystyle \sum_{\mu'} d_{\mu'} 
						\hat{B}_{\mu'}\hat{B}^\dagger_{\mu'} + \frac{1}{2}\LowTrace{H}
			\end{array}
			,\end{equation}
		which corresponds to Eq. (\ref{eq:FinalForm}) with $c = \frac{1}{2}\LowTrace{H}$.
		Also observe that in this way $d_j \leq 0$ and $d_{j+N} \geq 0$ for $j = 1,...,N$,
		which implies $d_{N+j}-d_{j} \geq 0$: we are satisfying also the Requirement 3. 
	\end{proof}

	Note that the vectors $\mathbf{y}_j$ can be normalized, if this is convenient, 
	and their order in the construction of a suitable $\tilde{R}$ function can be chosen arbitrarily.

	It is also important to observe that the coefficient matrix of a Hermitian operator
	can be non-Hermitian as well. 
	As an example consider
		\begin{equation}
			H' = H + i\Omega 
		.\end{equation}
	Noting that $\Omega_\Std = 0$ and $\LowTrace{\Omega} = 0$, and 
	exploiting the same reasoning that leads to Eq. (\ref{eq:Decomposition}), we obtain
		\begin{equation}
			\LowTrace{\Omega\hat{G}} = \LowTrace{\Omega_\Std \hat{G}} + \frac{1}{2} \LowTrace{\Omega}  = 0
		.\end{equation}
	Thus we have 
		\begin{equation}
			\hat{H} = \LowTrace{H'\hat{G}} \equiv \LowTrace{H\hat{G}}
		.\end{equation}

	As a final remark, observe that $H_\Std$ actually has the form usually 
	assumed in literature for the fermionic coefficient matrix
			\begin{equation}
				H_\Std = \Tonde{ 
					\begin{array}{cc} 
						E & F		\\
						-F^* & -E^*  	\\
					\end{array}
				}
				\label{eq:StandardFormLook}
			\end{equation}
	where the blocks $E$ and $F$ are Hermitian and anti-symmetric, respectively.

	\section{Numerical example}

	Consider a fermionic system with $N=2$, characterized by the coefficient matrix
		\begin{equation}
			\begin{array}{rl}
				H =& \Tonde{ \begin{array}{cccc}
					1	& 0	&	0	&	0	\\
					0	& 1	&	-i-1	&	0  	\\
					0	& i-1	&	1	&	-i-1	\\
					0	& 0	&	i-1	&	1  
				\end{array} } \\\\
			\end{array}
		.\end{equation}
	Clearly $H$ is not in the usual form shown in Eq. (\ref{eq:StandardFormLook}). 
	In the following it is shown, step by step, how to perform the transformation.
	\begin{enumerate}
	\item 
		Decompose $\hat{H}$ according to Eq. (\ref{eq:Decomposition}):
			\begin{equation}	
				H_\Std = \frac{1}{2}\Tonde{ \begin{array}{cccc}
					0	&	1-i	&	0	&	1+i	\\
					1+i	&	0 	&	-1-i     &	0     	\\
    					0       &	-1+i     &	0   	&	-1-i	\\
					1-i	&	0	&	-1+i	&	0    
				\end{array}}
			\end{equation}
		and
			\begin{equation}
				c = \frac{1}{2} \LowTrace{ H } = 2
			.\end{equation}
		Although the matrix $H$ is invertible, $H_\Std$ is singular, thus the special 
		treatment for singular coefficient matrices has to be used. 
	\item
		Calculate $\mathcal{N}_{H_\Std}$ and $\mathcal{K}_{H_\Std}$. 
		Possible choices are 

		\begin{equation}
			\mathcal{N}_{H_\Std} = \Graffe{ 
				\mathbf{x}_1 = 
					\frac{1}{2}
					\Tonde {  
					\begin{array}{c} 
						-\frac{1+i}{\sqrt{2}}   \vspace{0.1cm}\\ 
						i 			\vspace{0.1cm}\\ 
						\frac{1+i}{\sqrt{2}}    \vspace{0.1cm}\\ 
						1 
					\end{array}} }
		,\end{equation}
		\begin{equation}
			\mathcal{K}_{H_\Std} = 
				\Graffe{ 
				\mathbf{k}_1 = \Tonde{ 
					\begin{array}{c}  
						0			\vspace{0.1cm}\\
						\frac{1+i}{\sqrt{2}}	\vspace{0.1cm}\\
						0 			\vspace{0.1cm}\\
						\frac{-1+i}{\sqrt{2}}
					\end{array}
				}
				,\,
				\mathbf{k}_2 = \Tonde{
					\begin{array}{c}
						\frac{1}{\sqrt{2}} 	\vspace{0.1cm}\\
						0 		   	\vspace{0.1cm}\\
						\frac{1}{\sqrt{2}} 	\vspace{0.1cm}\\
						0
					\end{array}}
				}
		.\end{equation}

		Note here that $J(\mathbf{x_1})$ is orthogonal to $\mathbf{x}_1$, $\mathbf{k}_1$, and $\mathbf{k}_2$, 
		but $J(\mathbf{k}_1) \not\perp \mathbf{k_1}$ and $J(\mathbf{k}_2) \not\perp \mathbf{k_2}$. 

		At this stage we already know the final diagonal form of the Hamiltonian operator:
		the $d_\mu$ associated to $H_\Std$ are 
		$d_1 = -\sqrt{2}$, $d_2 = 0$, $d_3 = \sqrt{2}$, and $d_4=0$. Thus, we obtain 
			\begin{equation}
				\hat{H} = ( d_3 - d_1 )\ \hat{b}^\dagger_1\hat{b}_1  + d_1 + c = 
				2\sqrt{2}\ \hat{b}^\dagger_1\hat{b}_1 - \sqrt{2} + 2 
				\label{eq:ExampleSolution}
			.\end{equation}
	\item
		Build an orthogonal  J-invariant basis $\Graffe{ \mathbf{y}_1, \mathbf{y}_2 }$ for $\mathbb{K}$.
		We set
			\begin{equation}
				\mathbf{y}_1 = L(\mathbf{k}_1) = 
				\frac{i}{\sqrt{2}} \Tonde { \begin{array}{c}  0 \\ 1+i \\ 0 \\ -1+i \end{array}}
			\end{equation}
		and we look for a vector in $\mathbb{K}$ that is orthogonal to $\mathbf{y}_1$. Fortunately, 
		$\mathbf{k}_2$ is already orthogonal to $\mathbf{y}_1$ and it is also J-invariant:
		hence in this particular case 
			\begin{equation}
				\mathbf{y}_2 = \mathbf{k}_2 = \frac{1}{\sqrt{2}}\Tonde{ 
					\begin{array}{c}  1 \\ 0 \\ 1 \\ 0 \end{array}} 
			.\end{equation}
	\item
		Proceed into the identification of a $\tilde{R}$. 
		Let
			\begin{equation}
				\tilde{R} = i \Tonde{\Enlarge
						 \mathbf{y}_1  \Scalar{ \mathbf{y}_2 }{\cdot}
						 - \mathbf{y}_2  \Scalar{ \mathbf{y}_1 }{\cdot} }
			.\end{equation}
		Find its eigenvectors with strictly negative eigenvalue:
		its associated matrix in the canonical basis is
		\begin{equation}
			R = -\frac{1}{2}\Tonde{ \begin{array}{cccc}
				0	&	1-i	&	0	&	-1-i	\\
				1+i	&	0 	&	1+i 	&	0     	\\
    				0       &	1-i  	&	0   	&	-1-i 	\\
				-1+i	&	0	&	-1+i	&	0    
			\end{array}}
		,\end{equation}
		from which 
		\begin{equation}
			\mathcal{N}_R = \Graffe{ 
				\mathbf{k}'_1 =\frac{1}{2} 
			\Tonde{  \begin{array}{c} 
					-\frac{1+i}{\sqrt{2}}	\vspace{0.1cm}\\
					-i 			\vspace{0.1cm}\\
					-\frac{1+i}{\sqrt{2}}	\vspace{0.1cm}\\
					1 
			\end{array}} }
		.\end{equation}
		$\mathbf{k}'_1$ is in the kernel of $H_\Std$ and it is orthogonal to $J(\mathbf{k}'_1)$.
	\item
		\newcommand{\Frac}[2]{{\displaystyle \frac{#1}{#2}}}
		Finally, state that a valid $U$ is given by
			\begin{equation}\begin{array}{rl}
				\hspace{-1cm}
				U  =& \Tonde{\displaystyle\Enlarge \mathbf{x}_1, \mathbf{k}'_1, J(\mathbf{x}_1), J(\mathbf{k}_1') } \\\\
				=& {\displaystyle \frac{1}{2}}\Tonde{ 
					\resizebox{0.6\linewidth}{!}{$
					\begin{array}{cccc} 
	-\Frac{1+i}{\sqrt{2}}	&	-\Frac{1+i}{\sqrt{2}} &	\Frac{1-i}{\sqrt{2}} &	-\Frac{1-i}{\sqrt{2}}	\\\\
		i		&	-i     		&	1 			&	1		\\\\
    	\Frac{1+i}{\sqrt{2}}	&	-\Frac{1+i}{\sqrt{2}}	& -\Frac{1-i}{\sqrt{2}}	& -\Frac{1-i}{\sqrt{2}} \\\\
			1	&	1   		&	-i			&	i
			\end{array}$}} 
			\end{array}
			.\end{equation}
		This provides the connection between the initial $\hat{A}_\mu$ and the final $\hat{B_\nu}$ operators
		through the relationship in Eq. (\ref{eq:BAConnection}), and leads to 
		the diagonal form that we have written down in Eq. (\ref{eq:ExampleSolution}).
	\end{enumerate}

	\section{Conclusion}
			In this paper a concise and accessible treatment of
	the Bogoliubov-Valatin transformation for fermionic Hamiltonians
	without a linear part, which is the most common situation, has been provided.
	The procedure has been discussed making minimal assumptions and keeping the
	mathematical formalism as simple as possible. 
	The focus has been kept on some details that are somewhat overlooked in
	non-specialized literature, such as the properties of the coefficient matrix
	and the special treatment necessary in the singular matrix case. 
	A step-by-step numerical example completes this paper and guides the reader
	through the passages of the transformation.  For these reasons this work may serve both as 
	quick reference for the non-specialized researcher 
	and as supplementary material for a graduate-level course. 

	\appendix

	\section{Acknowledgments}
		I am deeply indebted to Prof. Davide Rossini for his encouragement and
		precious advice.

	\bibliography{refs}
	\bibliographystyle{unsrt}

	\end{multicols}

\end{document}